 \documentclass[preprint2]{aastex}

\newcommand{\myrule}{\rule[-0.1cm]{0.cm}{0.7cm}} 
\newcommand\msun{M_{\odot}}

\newcommand\mjup{M_\mathrm{Jup}}

\newcommand\chaha{Cha\,H$\alpha$\,}

\slugcomment{To appear in ApJ Letter}

\shorttitle{16--20\,$\mjup$ RV companion orbiting the brown dwarf candidate \chaha8}
\shortauthors{Joergens \& M\"uller}

\begin{document}


\title{
16--20\,$\mjup$ RV companion \\ orbiting the brown dwarf candidate \chaha8\thanks{Based 
on observations obtained at the Very Large Telescope of the 
European Southern Observatory at Paranal, Chile 
in program 75.C-0851(C), 77.C-0831(A+D), 278.C-5061(A).}}


\author{V. Joergens\altaffilmark{ } and
A. M\"uller\altaffilmark{ }}
\affil{Max-Planck Institute for Astronomy, 
     K\"onigstuhl~17, D-69117 Heidelberg, Germany
\email{viki@mpia.de, amueller@mpia.de}
accepted for publication in ApJ Letter
}



\begin{abstract}
We report the discovery of a 16--20\,$\mjup$ radial velocity companion around the 
very young ($\sim$3\,Myr) brown dwarf candidate \chaha8 (M5.75--M6.5).
Based on high-resolution echelle spectra 
of \chaha8 taken between 2000 and 2007 
with UVES at the VLT, a companion was detected through RV variability with a 
semi-amplitude of 1.6\,km\,s$^{-1}$. A Kepler fit to the data 
yields an orbital period of the companion of 1590~days and an eccentricity of $e$=0.49. 
A companion minimum mass $M_2\sin i$ between 16 and 20\,$\mjup$ is derived when
using model-dependent mass estimates for the primary.
The mass ratio $q\equiv M_2/M_1$ might be as small as 0.2 and, with 
a probability of 87\%, it is less than 0.4.
\chaha8 harbors most certainly the lowest mass companion detected so far
in a close ($\sim$ 1\,AU) orbit around a brown dwarf or very low-mass star. 
From the uncertainty in the orbit solution, it 
cannot completely be ruled out that the companion has a mass in the planetary 
regime. 
Its discovery is in any case an important step towards RV planet detections around BDs.
Further, 
\chaha8 is the fourth known spectroscopic brown dwarf or very low-mass binary system with an
RV orbit solution and the second known very young one. 
\end{abstract}





\section{Introduction}

Search for planetary or brown dwarf (BD) companions to BDs
are of primary interest for understanding planet and BD formation.
There exists 
no widely accepted model for the formation of BDs 
(e.g. recent review by Luhman et al. 2007).
The frequency of BDs in multiple systems is a fundamental parameter
in these models.
However, it is poorly constrained for close separations: 
Most of the current surveys for companions
to BDs are done by direct (adaptive optics or HST) imaging and are not sensitive to close 
binaries ($a\lesssim1$\,AU and $a\lesssim10$\,AU for the field and clusters, 
respectively),
and found preferentially close to equal mass systems 
(e.g. Bouy et al. 2003; see also Burgasser et al. 2007). 
Spectroscopic monitoring for radial velocity (RV) variations provides a means to detect 
close systems.
The detection of the first spectroscopic BD binary in the Pleiades, PPl\,15 
(Basri \& Mart\'\i n 1999), raised hope to find many more of these systems in the 
following years. 
However, the number of confirmed close companions to BDs and very low-mass stars (VLMS,
$M \leq 0.1\,\msun$) 
is still small.
While several spectroscopic companions to BD/VLMS
were reported recently (e.g. Reid et al. 2002; Guenther \& Wuchterl 2003;
Kenyon et al. 2005; Kurosawa, Harries \& Littlefair 2006; 
Basri \& Reiners 2006; L. Prato in preparation),
many of these companion identifications 
are based on only 2--3 RV measurements for an individual target. 
To date, there are three spectroscopic BD binaries confirmed, i.e. for which a spectroscopic
orbital solution has been derived: the before mentioned double-lined spectroscopic
binary (SB2) PPl\,15, the very young eclipsing SB2 system in Orion 
2MASS~J05352184-0546085 
(2M0535-05, Stassun, Mathieu \& Valenti 2006), and a SB2 within the quadruple 
GJ\,569 (Zapatero Osorio et al. 2004; Simon, Bender \& Prato 2006). 
All these systems have a mass ratio close to unity. In particular, no RV planet
of a BD/VLMS has been found yet.
If BDs can harbor planets at a few AU distance is still unknown.
Among the more than 200 extrasolar planets that have been detected around stars
by the RV technique, 6 orbit stellar M-dwarfs 
(e.g. Udry et al. 2007) showing that planets can form also around 
primaries of substantially lower mass than our Sun.
Observations hint that basic ingredients for planet formation
(disk material, grain growth) 
are present also for BDs (e.g. Apai et al. 2005).
However, the only planet detection around
a BD is a very wide 55\,AU system (2M1207, Chauvin et al. 2005; 
cf. also Caballero et al. 2006 for another candidate), which
is presumably formed very differently from
the Solar System and RV planets.

RV surveys for planets around such
faint objects, as BD/VLMS are, require monitoring with high spectral dispersion
at 8--10\,m class telescopes. While being expensive in terms of telescope
time, this is, nevertheless, extremely important for our understanding
of planet and BD formation.
Within the course of an RV survey for (planetary and BD) companions to very young 
BD/VLMS in Cha\,I
(Joergens \& Guenther 2001; Joergens 2006), 
evidence for a very low-mass
companion orbiting the BD candidate \chaha8 was found (Joergens 2005, 2006).
We report here on follow-up RV monitoring, which confirms the companion and, combined
with previous RV measurements, allows to determine an RV orbit.

\section{The host object \chaha8}
\label{sect:host}

Cha\,H$\alpha$\,8\footnote{Simbad name: [NC98]~Cha~HA~8} has been identified as very low-mass member of the nearby
Chamaeleon\,I star-forming region ($\sim$160\,pc) by an H$\alpha$ objective prism survey
and low- and medium-resolution spectroscopy 
(Comer\'on, Rieke \& Neuh\"auser 1999; Comer\'on, Neuh\"auser \& Kaas 2000;
Neuh\"auser \& Comer\'on 1998, 1999).
Membership in the Cha\,I association and, therefore, 
the youth of \chaha8, is well established based on 
H$\alpha$ and X-ray emission, lithium absorption, and RVs 
(see references above; Joergens \& Guenther 2001; Stelzer, Micela \& Neuh\"auser 2004; 
Joergens 2006).
Its spectral type has been determined to be between M5.75 (Luhman 2004, 2007) and 
M6.5 (Comer\'on et al. 2000).
Comer\'on et al. (2000) estimate a mass of
0.07\,$\msun$ and an age of 3\,Myr by employing evolutionary models by Baraffe et al. (1998).
Using slightly different values for effective temperature and bolometric luminosity 
by Luhman (2007), a mass of 
0.10\,$\msun$ and an age of 2.5\,Myr are found by comparison with the same models. 
Thus, \chaha8 is either a BD or a VLMS. 
While for many of the known substellar objects in Cha\,I circumstellar disks were detected
through mid-IR (Persi et al. 2000; Comer\'on et al. 2000) 
and $L$-band (Jayawardhana et al. 2003) excess emission,
no indications for disk material has been found for \chaha8 in these works.
See Table\,\ref{tab:hostparam} for a list of properties of \chaha8.


\section{Radial velocities and orbital solution}

Spectroscopic observations of \chaha8 were carried out between 2000 and 2007 with
the Uv-Visual Echelle Spectrograph (UVES) attached to the VLT 8.2\,m KUEYEN telescope
at a spectral resolution $\lambda$/$\Delta \lambda$ of 40\,000 in the 
red optical wavelength regime.
RVs were measured from these spectra based on a cross-correlation technique
employing telluric lines for the wavelength calibration. 
The errors of the relative RVs of \chaha8 range between 30 and 500\,m/s.
Details on the data analysis can be found in Joergens (2006).

\begin{figure*}[t]
\centering
\includegraphics[width=0.8\linewidth,clip]{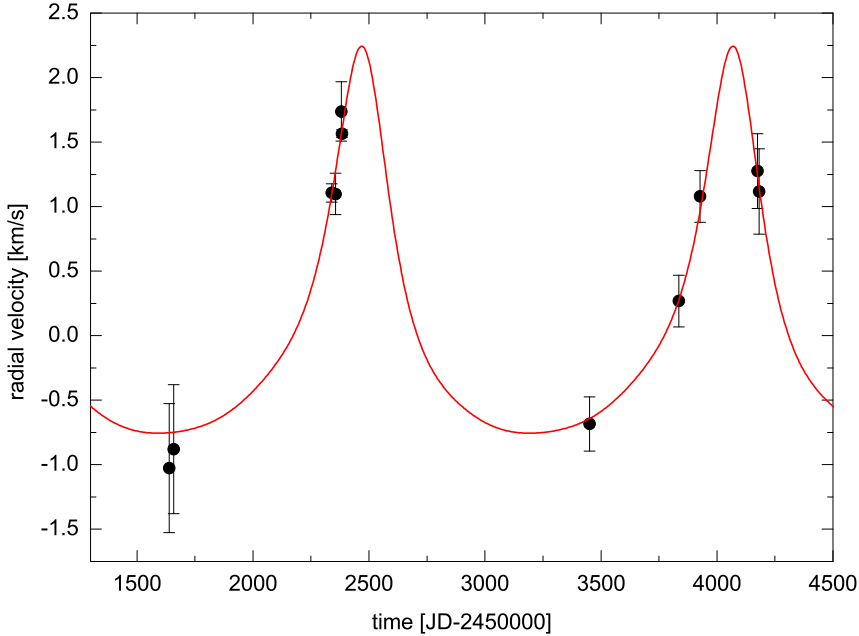} 
\caption{
\label{fig:orbit}
RV measurements of \chaha8 between 2000 and 2007 based on UVES/VLT spectra.
Overplotted is the best-fit Keplerian orbit, which has a semi-amplitude of 
1.6\,km\,s$^{-1}$, a period of 4.4\,years and an eccentricity of $e$=0.49.
}
\end{figure*}

RV measurements from spectra taken between 2000 and 2004 
indicated already the presence of an RV companion to \chaha8
(Joergens 2005, 2006).
This paper reports on follow-up RV monitoring between March 2005 and March 2007.
The new RV measurements are presented in Table\,\ref{tab:rvs}. Using the combined
RV data from 2000 to 2007, it was possible to derive a
spectroscopic orbital solution.
Fig.\,\ref{fig:orbit} shows the RV measurements
together with the RV curve of the best-fit Kepler model.
The reduced $\chi^2$ of the orbital fit is 0.42.
The fitted Kepler orbit is that of a companion with a mass function of 
4.6 $\times 10^{-4}\,\msun$ revolving \chaha8 with a period of 1590\,d (4.4\,yr) 
on an eccentric ($e$=0.49) orbit and causing 
an RV semi-amplitude of 1.6$\pm$0.4\,km\,s$^{-1}$. The semi-major axis is of the order of
1\,AU. See Table\,\ref{tab:orbitparam} for the whole set of 
orbital elements.

\section{Companion mass and system mass ratio}

The small RV semi-amplitude of \chaha8 and the fact that
no spectral lines of the companion were detected at any orbital phase hints
already at a small companion mass.
The mass of the companion $M_2 \sin i$ cannot be determined directly
from a single-lined RV orbit but depends on the primary mass.
Unfortunately, in the case of \chaha8, the primary mass is not very precisely
determined (as common in this mass and age regime). Using
the two available estimates for the primary mass
(0.07 and 0.10\,$\msun$, see Sect.\,\ref{sect:host}),  
the mass of the companion $M_2 \sin i$ is inferred to
15.6 and 19.5$\mjup$, respectively.
This does not take into account further possible errors in the primary mass,
as e.g. introduced by evolutionary models.
Given the uncertainty of the RV semi-amplitude (0.4\,km\,s$^{-1}$) of the orbit solution, 
it cannot be ruled out that the companion 
has a mass $M_2 \sin i$ in the planetary mass
regime ($<13\,\mjup$). The reason for this is the limited phase coverage of the 
available RV data, in particular at maximum RV. 
For example, for an RV semi-amplitude of 1.4\,km\,s$^{-1}$, 
the companion mass $M_2 \sin i$ would be 11.6\,$\mjup$ ($M_1=0.07\,\msun$).

Based on the assumption of randomly oriented orbits in space, the following 
statements about the mass ratio of the system can be made: 
With a 50\% probability (inclination $i \geq$ 60\,deg), the mass ratio q$\equiv$M$_2$/M$_1$
is $\lesssim$0.2, and with 87\% probability ($i \geq$ 30\,deg), q is $\lesssim$0.4.
Comparing the mass ratio of \chaha8 with that of other BD/VLM spectroscopic binaries
(q$>$0.6, Basri \& Mart\'\i n 1999; Stassun et al. 2006; Simon et al. 2006), 
with a probability of more than 90\%, \chaha8 has the smallest known
mass ratio (q$\lesssim$0.5). 
It is noted that these probabilities are valid for both considered values of M$_1$.
Further, these are the probabilities for randomly oriented orbits. They are
even higher for spectroscopic systems since this search method has a bias towards high
inclinations. 


\section{Activity}

In the following, the question is addressed whether the detected RV variability 
with semi-amplitude of 1.6\,km\,s$^{-1}$ can be caused by chromospheric or accretion
activity, which are common phenomena for very young stars and are observed 
also for substellar objects. 
\chaha8 shows signs for chromospheric activity through H$\alpha$ 
(Comer\'on et al. 2000;  Luhman 2004; Mohanty et al. 2005)
and X-ray emission (Stelzer et al. 2004).
Chromospheric activity can cause photometric and RV variability on the time
scale of the rotation period through asymmetries in the surface brightness
and spectral line shape, respectively.
The rotation period of \chaha8 is of the order of a few days based on 
measurement of its spectroscopic velocity $v \sin i$ (Joergens \& Guenther 2001,
cf. Table\,\ref{tab:hostparam})
and in accordance with absolute rotation periods determined for BD/VLMS 
in the same region ($P_{rot}$=2--5\,days, Joergens et al. 2003).
Photometric and RV variability of \chaha8 on this time scale is of rather low amplitude:
Photometric monitoring in the Bessel $R$ and Gunn $i$ filter (Joergens et al. 2003) 
over 6 nights show that peak-to-peak variability amplitudes are
$\Delta R<$0.02\,mag and $\Delta i<$0.04\,mag, respectively.
Further, while we report here on RV variability with a period of a few years,
RV variability of \chaha8 on time scales of days is quite small: 
Investigation of RVs measured with time offsets of a few days in 2000 
($\Delta t$=19\,d),
in March 2002 (16\,d), April 2002 (3\,d) and in 2007 (7\,d) show that 
peak-to-peak RV differences
on time scales of a few days do not exceed 0.17\,km\,s$^{-1}$ and can account for 
only about a tenth of the total recorded variability amplitude.
Therefore, the detected long-period
RV variability cannot be explained by rotational modulation due to
chromospheric activity.
Accretion, on the other hand, can cause RV variability on various time scales.
However, since
no signs for significant accretion were detected for \chaha8
(H$\alpha$ equivalent width $\leq$10\,{\AA},
no Ca\,II $\lambda$ 8662\,{\AA} emission detected; Mohanty et al. 2005), 
accretion processes are unlikely to cause the detected RV variability.

\section{Discussion and Conclusions}

We have shown that \chaha8 has an RV companion in a 
$\sim$1\,AU orbit and that this companion is most certainly a very low-mass BD.
\chaha8 is the first small mass ratio spectroscopic binary
among BD/VLMS. 
The discovery of the RV companion of \chaha8, 
which has an RV semi-amplitude of only 1.6\,km\,s$^{-1}$,
is an important step towards RV planet detections of BD/VLMS.
In fact, from the uncertainty in the orbit solution, it 
cannot be completely excluded that the companion of \chaha8
has a mass in the planetary mass regime ($<13\,\mjup$). 
Follow-up RV measurements monitoring the next phase of periastron
(April 2011) are necessary to investigate this further.

The favored mechanisms for stellar binary formation, fragmentation of collapsing 
cloud cores
or of massive circumstellar disks, seem to produce preferentially equal mass
components, in particular for close separations
(e.g. Bate et al. 2003).   
Thus, they have difficulties to explain the formation of
the small mass ratio system \chaha8. However,
we know that close \emph{stellar} binaries with small mass ratios do exist as well
(e.g. q=0.2, Prato et al. 2002), and without knowing the exact mechanism
by which they form, it might be also an option for \chaha8. 

Considering the small mass of the companion of \chaha8, 
a planet-like formation could also be possible.
Giant planet formation through core accretion might be hampered
for low-mass primaries, like M dwarfs, by long formation time scales 
(Laughlin, Bodenheimer \& Adams 2004; Ida \& Lin 2005), 
though, recent simulations hint that it can be
a faster process than previously anticipated (Alibert et al. 2005).
On the other hand, giant planets around M dwarfs might form by disk instability
(Boss 2006a, 2006b), at least in low-mass star-forming regions,
where there is no photoevaporation of the disk through nearby hot stars (e.g. Cha\,I).
The companion of \chaha8 could have been formed through disk instability,
either in situ at 1\,AU or, alternatively, at a larger separation and subsequent
inwards migration. The latter is plausible in the case of \chaha8 since
a higher mass of the formed object favors
inward migration (e.g., Boss 2005).

\chaha8 outstands the group of spectroscopic BD/VLM binaries also by its 
relatively long orbital period (1590\,d). 
In particular, this is the case when considering
the extremely short period systems PPl\,15 and 2M0535-05, 
which have periods smaller than 10 days. \chaha8 is part of a sample 
of ten BD/VLMS in Cha\,I that have been monitored for RV companions since 2000.
Joergens (2006; V. Joergens in preparation) finds no short period
systems (and also no equal mass binaries) 
in this survey, while they are easier to detect.
Thus, the detection of \chaha8 might hint at a higher frequency of 
long-period ($\sim$10$^3$\,d) BD/VLM binaries in Cha\,I 
than short period ones ($\sim$10\,d).
This is consistent with the separation distribution for currently known substellar and
very low-mass stellar binaries
(Burgasser et al. 2007) which has a peak at 2.5--10\,AU. We note, however, that
this distribution is not well constrained 
for separations $<$3\,AU.



When combined with 
angular distance measurements or eclipse detections,
spectroscopic binaries allow valuable dynamical mass determinations.
The mass is the most important input parameter for evolutionary models, which rely
for masses $<$0.3\,M$_\mathrm{\odot}$, only on the two masses 
determined for the very young eclipsing BD binary 
2M0535-05 (Stassun et al. 2006; Mathieu et al 2007).  
\chaha8 is, after 2M0535-05, the second known very young BD/VLM 
spectroscopic binary. 
However, the measurement of dynamical masses for \chaha8 is challenging in several respects.
In order to measure absolute masses of both components, it is required to resolve the
spectral lines of both components (SB2). 
This is preferentially done at IR wavelengths (e.g. CRIRES/VLT), were the contrast ratio
between primary and secondary is smaller (e.g. Prato 2007). 
Having a maximum separation of about 13\,milli\,arcsec ($\sim$2\,AU), 
the spatial resolution of current imaging instruments
is not sufficient to directly resolve \chaha8.
This has probably to await OWL.
Current and upcoming interferometers, on the other hand, do provide the necessary spatial resolution,
but are not sensitive enough.
However, it might be possible to detect the relative astrometric signal of the primary 
(few milli\,arcsec). This would allow measurement
of the inclination of the orbital plane and, therefore, breaking the 
$\sin i$ ambiguity in the companion mass. 
These observations might be possible with, e.g. NACO/VLT or
with phase-referenced astrometry with the
upcoming VLT Interferometer PRIMA (using available brighter 
reference stars in the field for fringe tracking).

\acknowledgments
We are grateful to the director general of ESO for 
granting DDT observations of \chaha8. 
It is also a pleasure to acknowledge the excellent work of the 
ESO staff at Paranal, who took the UVES spectra of this paper
in service mode. 
Further, we would like to thank our colleagues
Johny Setiawan, Sabine Reffert, and Jose
Caballero for helpful discussions and comments.
This publication has made use of the VLM Binaries Archive maintained by 
Nick Siegler at http://paperclip.as.arizona.edu/$\sim$nsiegler/VLM\_binaries/. 



{\it Facilities:} \facility{ESO VLT (UVES)}.

\clearpage

\clearpage

\begin{table}
\begin{center}
\caption{
\label{tab:hostparam} 
Properties of \chaha8.
}
\begin{tabular}{llcc}
\tableline
\tableline
& & \\
\multicolumn{2}{l}{Parameter}  & \chaha8 & Reference \\
\tableline
\myrule
SpT           &               & M6.5, M5.75   & 1,2\\
V             & [mag]         & 20.1          & 1 \\
$T_\mathrm{eff}$  & [K]       & 2910, 3024    & 1,2\\
$log(L)$      & [$L_{\odot}$] & -1.65, -1.43  & 1,2\\
$M_1$         & [$\msun$]     & 0.07, 0.10     & 1,2 \\

$v \sin i $   & [km\,s$^{-1}$] & 15.5$\pm$2.6 & 3 \\
$P_{v \sin i}$ & [days]      & 1.9            & 3 \\
$\Delta R$ & [mag]         & $<$0.02          & 4 \\
$\Delta i$ & [mag]         & $<$0.04          & 4 \\
EW(H$_{\alpha}$) & [\AA]   & 9, 8.4, 10       & 1,5,6\\

\tableline
\end{tabular}
\tablerefs{
(1) Comer\'on et al. 2000;
(2) Luhman 2007;
(3) Joergens \& Guenther 2001; 
(4) Joergens et al. 2003;
(5) Mohanty et al. 2005;
(6) Luhman 2004.
}
\end{center}
\end{table}

\begin{table}
\begin{center}
\caption{
\label{tab:rvs} 
New RV measurements of Cha\,H$\alpha$\,8.
}
\begin{tabular}{lllll}
\tableline
\tableline
\myrule
Object               &   Date      & HJD           & RV     & ~$\sigma_{RV}$ \\
                     &             &       & [km\,s$^{-1}$] & [km\,s$^{-1}$] \\
\tableline
\myrule
Cha\,H$\alpha$\,8    & 2005 Mar 21 & 2453450.62080  & 15.130$^a$ & 0.21\, \\ 

                     & 2006 Apr 10 & 2453835.65109  & 16.082 & 0.20 \\ 
                     & 2006 Jul 09 & 2453926.50137  & 16.893 & 0.20 \\ 

                     & 2007 Mar 15 & 2454174.66101  & 17.089$^a$ & 0.29 \\ 
                     & 2007 Mar 22 & 2454181.69756  & 16.931$^a$ & 0.33 \\ 

\tableline

\end{tabular}
\tablecomments{HJD is given at the middle of the exposure. 
$\sigma_{RV}$ is the estimated error of the relative RVs.
An additional error of 400\,m\,s$^{-1}$ has to be
taken into account for the absolute RVs.
}
\tablenotetext{a}{RV value is the average of two single consecutive measurements. The corresponding error
$\sigma_{RV}$ is the standard deviation of the individual measurements.}
\end{center}
\end{table}
\clearpage
\begin{table}
\begin{center}
\caption{
\label{tab:orbitparam} 
Orbital and physical parameters derived for the best-fit Keplerian model
of Cha\,H$\alpha$\,8.
}
\begin{tabular}{llcc}
\tableline
\tableline
& & \\

& & \multicolumn{2}{c}{Cha\,H$\alpha$\,8} \\
\multicolumn{2}{l}{Parameter}  & Value & Error \\

\tableline
\myrule

P             & [days]             & 1590.9     &  21.1      \\
T             & [HJD-2450000]      & 2487.5     &  87.3      \\
e             &                    & 0.49       &   0.19     \\
V             & [km\,s$^{-1}$]     & 15.774     &   0.212    \\
$\omega$      & [deg]              & 8.20       &  $^{+40.2}_{-8.20}$\\
K             & [km\,s$^{-1}$]     & 1.615      &  0.366      \\
$a_1 \sin i$ & [AU]               & 0.21       & 0.10        \\
$f(m)$        & [$10^{-4}\,\msun$] & 4.599      &             \\
$M_2 \sin i $  & [$\mjup$]          & 15.6, 19.5 $^a$ & 0.6, 0.8 $^a$   \\
$a_2 $         & [AU]               & 0.97, 1.10 $^a$ & 0.10, 0
.13 $^a$ \\

\tableline

$N_{\rm meas}$        &        & 11    \\
$Span$            & [days] & 2542  \\ 
$\sigma$ (O-C)    & [m/s]  & 96.7  \\
$\chi^{2}_{\rm red}$  &        & 0.424  & \\

\tableline
\end{tabular}
\tablecomments{
The given parameters are: orbital period,  
periastron time, eccentricity, system velocity, longitude of periastron,
RV semi-amplitude, projected semi-major axis of the primary,
mass function, lower limit of the companion mass, 
semi-major axis of the companion, number of measurements,
time span of the observations, residuals, reduced $\chi^2$. 
}
\tablenotetext{a}{Derived parameter based on two available estimates 
for the primary mass of 0.07\,$\msun$ and 0.10\,$\msun$.
No further errors of the primary mass, e.g. as introduced by the use of
evolutionary models, have been taken into account here.}
\end{center}
\end{table}








\end{document}